\documentclass[12pt]{article}
\usepackage{amssymb}
\usepackage{amsfonts}
\usepackage{amsmath}
\usepackage{revsymb}
\usepackage{graphicx}

\begin{document}

\begin{flushright}
HIP-2009-10/TH
\end{flushright}

\vskip .7cm

\begin{center}
{\Large\bf Dirac Equation in Noncommutative Space\\ for Hydrogen
Atom\\\vskip0.3cm }

\vskip .7cm

{\bf\large T. C. Adorno$^1$, M. C. Baldiotti$^1$, M.
Chaichian$^2$,\\ D. M. Gitman
$^1$ and A. Tureanu$^2$}\\

\vskip .7cm

{\it $^1$Instituto de F\'{\i}sica, Universidade de S\~{a}o Paulo,
Caixa Postal
66318,\\ CEP 05508-090 S\~{a}o Paulo, S.P., Brazil\\
$^2$Department of Physics, University of Helsinki and Helsinki
Institute of Physics,\\ P.O.Box 64, FIN-00014 Helsinki, Finland}

\end{center}

\vskip1cm
\begin{abstract}
{We consider the energy levels of a hydrogen-like atom in the
framework of $\theta $-modified, due to space noncommutativity,
Dirac equation with Coulomb field. It is shown that on the
noncommutative (NC) space the degeneracy of the levels $2S_{1/2},\,
2P_{1/2}$ and $ 2P_{3/2}$ is lifted completely, such that new
transition channels are allowed.}
\end{abstract}

\section{Introduction}

Possible physically observable consequences of noncommutativity of
space coordinates have attracted recently a lot of attention in
order to justify the intensive study of noncommutative versions of
QFT and quantum mechanics, mainly motivated by the seminal works
\cite{DFR,SW}. The results of Ref. \cite{ChaiSheTur01} show, e.g.,
that the noncommutativity leads to a deviation of the energy levels
of the hydrogen atom from the well-known structure obtained in the
framework of standard quantum mechanics. The results of
\cite{ChaiSheTur01} show that the degenerate levels $2S-2P$ split in
three others levels. However, the
calculations of \cite{ChaiSheTur01} were done in the framework of $\theta$%
-modified nonrelativistic Schr\"{o}dinger equation with Coulomb
field. It is interesting to refine these results, considering a
noncommutative modification of
the energy levels of the hydrogen atom in the framework of the $\theta$%
-modified Dirac equation with the Coulomb field, which is the
relativistic Schr\"{o}dinger equation for the case under
consideration. Such a problem is solved in the present work. We show
that in contrast with the
nonrelativistic case, the noncommutativity lifts completely the degeneracy of the levels $%
2P_{1/2}$ and $2P_{3/2}$, opening new allowed transition channels.

\section{$\protect\theta$-modified Dirac equation with Coulomb field}

The behavior of an electron with charge $-e$ $\left( e>0\right) $
and mass $m $ in the Coulomb field of a nucleus\emph{\ }$Ze$, in a
noncommutative space, is determined by the $\theta $-modified Dirac
equation, whose Hamiltonian form
is%
\begin{align}
& i\frac{\partial }{\partial t}\Psi \left( x\right) =\hat{H}^{\theta }\Psi
\left( x\right) \,,\ \ \hat{H}^{\theta }=\left( \boldsymbol{\alpha }\cdot
\mathbf{\hat{p}}\right) +m\gamma ^{0}-eA_{0}\left( q\right) \,,  \notag \\
& \alpha _{i}\equiv \gamma _{0}\gamma _{i}\,,\ \ \boldsymbol{\alpha }=\left(
\alpha _{i}\right) \,,\ \ \mathbf{\hat{p}}=\left( p_{i}\right) \,,
\label{1.1}
\end{align}%
where $\Psi \left( x\right) $ is a four component spinor, $\mathbf{\hat{p}}%
=-i\mathbf{\nabla }$, $\gamma _{\mu }$ are the Dirac matrices and the $%
\theta $ -modified Coulomb potential $A_{0}\left( q\right) $ reads
\begin{equation}
A_{0}\left( q\right) =A_{0}\left( x^{i}-\frac{1}{2}\theta ^{ij}\hat{p}%
_{j}\right) =\frac{Ze}{\sqrt{\left( x^{i}-\frac{1}{2}\theta ^{ij}\hat{p}%
_{j}\right) \left( x^{i}-\frac{1}{2}\theta ^{ik}\hat{p}_{k}\right) }}\,,\ \
\ i,j,k=(1,2,3)\,,  \label{1.3}
\end{equation}%
where $\theta ^{jk}$ is the constant anti-symmetric noncommutativity
parameter that
defines the algebra of noncommutative position operators $q^{i}$, $\left[ q^{j},q^{k}%
\right] =i\theta ^{jk}$ (see, e.g., \cite{GitKu07}). In what follows we call
$\hat{H}^{\theta }$ the $\theta $-modified Dirac Hamiltonian with Coulomb
field on a noncommutative space. Setting $\theta ^{i}=\varepsilon _{ijk}\theta ^{jk}$ ($%
\varepsilon _{ijk}$ is the Levi-Civita simbol with the usual definition) and
denoting $r=\sqrt{x^{i}x^{i}}$, the potential (\ref{1.3}) can be written as

\begin{equation}
A_{0}\left( x^{i}-\frac{1}{2}\theta ^{ij}\hat{p}_{j}\right) =\frac{Ze}{r}+%
\frac{1}{4}\frac{Ze}{r^{3}}\left( \mathbf{\hat{L}}\cdot \boldsymbol{\theta }%
\right) +O\left( \theta ^{2}\right) \,\mathbf{,\ \ }\boldsymbol{\theta }%
=\left( \theta ^{i}\right) \,,  \notag
\end{equation}%
where $\mathbf{\hat{L}=}\left[ \mathbf{x}\times \mathbf{\hat{p}}\right] $ is
the angular momentum operator. Therefore the $\theta $-modified Dirac
Hamiltonian takes the form%
\begin{align}
& \hat{H}^{\theta }=\hat{H}+\hat{V}^{\theta }+O\left( \theta ^{2}\right) \,,
\label{1-12} \\
& \hat{H}=\left( \boldsymbol{\alpha }\cdot \mathbf{p}\right) +m\gamma ^{0}-%
\frac{Ze^{2}}{r}\,,\ \ \hat{V}^{\theta }=-\frac{1}{4}\frac{Ze^{2}}{r^{3}}%
\left( \mathbf{\hat{L}}\cdot \boldsymbol{\theta }\right) \,.  \label{1.2}
\end{align}

Using the exact eigenfunctions of $\hat{H}$ and treating $\hat{V}^{\theta }$
as a perturbation of the ordinary Dirac Hamiltonian $\hat{H}$, one can
calculate the modification of energy levels of the hydrogen atom in the
framework of $\theta $-modified Dirac equation. The spectrum of $\hat{H}$
and the corresponding eigenfunctions are well known, see \cite%
{Bethe,Rose,Akhiezer,VorGiT07}. Below, we list the spectrum and the
eigenfunctions of $\hat{H}$\ for $Z<137$:%
\begin{align}
& \hat{H}\Psi _{N,j,M,\zeta }=E_{N,j}\Psi _{N,j,M,\zeta }\,,\ \ \zeta =\pm
1\,,\ \ E_{N,j}=\frac{m}{\sqrt{1+\left( \frac{Ze^{2}}{\gamma +N}\right) ^{2}}%
}\,,  \notag \\
& N=\left\{
\begin{array}{l}
1,2,...\,,\ \ \zeta =1 \\
0,1,2...\,,\ \ \zeta =-1%
\end{array}%
\right. \,,\ \ \gamma =\sqrt{\varkappa ^{2}-\left( Ze^{2}\right) ^{2}}\,,\ \
\varkappa =\zeta \left( j+\frac{1}{2}\right) \,,  \notag \\
& \mathbf{\hat{J}}^{2}\Psi _{N,j,M,\zeta }=j\left( j+1\right) \Psi
_{N,j,M,\zeta }\,,\ \ \hat{J}_{3}\Psi _{N,j,M,\zeta }=M\Psi _{N,j,M,\zeta
}\,,  \notag \\
& \mathbf{\hat{J}=\mathbf{\hat{L}}}+\frac{1}{2}\boldsymbol{\Sigma \,}\mathbf{%
\mathbf{,\ \ }}\boldsymbol{\Sigma }=\mathrm{diag}\left( \boldsymbol{\sigma },%
\boldsymbol{\sigma }\right) \,,\ \ j=\frac{1}{2},\frac{3}{2},\ldots \,,\ \
-j\leq M\leq j\,,  \notag
\end{align}%
where $\boldsymbol{\sigma }=\left( \sigma _{i}\right) $ are the Pauli
matrices and the bispinors $\Psi _{N,j,M,\zeta }$ are given by%
\begin{align}
& \Psi _{N,j,M,\zeta }\left( r,\vartheta ,\varphi \right) =\frac{1}{r}\left(
\begin{array}{c}
\Omega _{j,M,\zeta }\left( \vartheta ,\varphi \right) F_{N,j,\zeta
}^{+}\left( r\right)  \\
i\Omega _{j,M,-\zeta }\left( \vartheta ,\varphi \right) F_{N,j,\zeta
}^{-}\left( r\right)
\end{array}%
\right) \,,  \notag \\
& \Omega _{j,M,\zeta }\left( \vartheta ,\varphi \right) =\frac{\zeta }{\sqrt{%
2j+\zeta +1}}\left(
\begin{array}{c}
-\sqrt{\left( j+\frac{\zeta }{2}\right) -\zeta M+\frac{1}{2}}Y_{j+\frac{%
\zeta }{2},M-\frac{1}{2}}\left( \vartheta ,\varphi \right)  \\
\sqrt{\left( j+\frac{\zeta }{2}\right) +\zeta M+\frac{1}{2}}Y_{j+\frac{\zeta
}{2},M+\frac{1}{2}}\left( \vartheta ,\varphi \right)
\end{array}%
\right) \,,  \label{1.12}
\end{align}%
where $Y_{A,B}\left( \vartheta ,\varphi \right) $ are the spherical
harmonics \cite{Rose,Akhiezer}\ and%
\begin{align}
& F_{N,j,\zeta }^{\pm }\left( r\right) =\pm A\lambda _{N}^{\pm }z^{\gamma
-1}e^{-z/2}\left[ \left( \eta _{N}-\varkappa \right) \Phi \left( -N,\beta
;z\right) \mp N\Phi \left( -N+1,\beta ;z\right) \right] \,,  \notag \\
& A=\frac{\left( 2\lambda _{N}\right) ^{3/2}}{2\Gamma \left( \beta \right) }%
\sqrt{\frac{\Gamma \left( \beta +N\right) }{\eta _{N}\left( \eta
_{N}-\varkappa \right) N!}}\,,\ \ \lambda _{N}^{\pm }=\sqrt{1\pm \frac{%
E_{N,j}}{m}}~,\ \lambda _{N}=m\lambda _{N}^{+}\lambda _{N}^{-}\,,  \notag \\
& \beta =2\gamma +1\,,\ \ \eta _{N}=Ze^{2}m/\lambda _{N}\,,\ \ z=2\lambda
_{N}r\,,  \label{1.13}
\end{align}%
where $\Phi \left( a,b;z\right) $ is the hypergeometric confluent function
(see \cite{Gradshteyn} for definition).

\section{$\protect\theta$-modification of energy levels}

As it was shown in \cite{ChaiSheTur01}, for the hydrogenic atoms, without
spin-orbit interaction, the nonrelativistic energy levels of the $\theta $%
-modified Hamiltonian $\hat{H}^{\mathrm{nr}}$,%
\begin{equation}
\hat{H}^{\mathrm{nr}}=\frac{\mathbf{\hat{p}}^{2}}{2m}-\frac{Ze^{2}}{r}\,,
\label{1.26b}
\end{equation}%
are characterized by a lift of degeneracy in the quantum number $l=j+\zeta /2
$ (eigenvalue of $\mathbf{\hat{L}}^{2}$ operator), implying certain
splittings of the energy levels. Thus, there appears a possibility of new
transition channels with distinct projections of total angular momentum $M$,
i.e., in the spectroscopic notation, transitions like $nl_{j}^{M}%
\longrightarrow nl_{j}^{M^{\prime }}$ (for $l\neq 0$), where
$n=N+\left\vert \varkappa \right\vert $ is the nonrelativistic
principal quantum number, eigenvalue of the radial differential
operator of (\ref{1.26b}). The perturbation due to the
noncommutativity is given by the same operator $\hat{V}^{\theta }$
as in (\ref{1.2}), but it is calculated with the nonrelativistic
wave functions, whose correction on degenerate levels, in leading
order, is
obtained by computing the eigenvalues of the secular matrix $\Delta E^{%
\mathrm{nr}}$, its elements being defined as%
\begin{equation}
\Delta E_{aa^{\prime }}^{\mathrm{nr}}=-\frac{Ze^{2}}{4}\int_{0}^{\infty
}dr\,r^{2}\int_{0}^{4\pi }d\Omega \left\{ \psi _{n,a}^{\dagger }\left(
r,\vartheta ,\varphi \right) \left[ \frac{\mathbf{\hat{L}}\cdot \boldsymbol{%
\theta }}{r^{3}}\right] \psi _{n,a^{\prime }}\left( r,\vartheta ,\varphi
\right) \right\} \,,  \label{1.26a}
\end{equation}%
where the label $a=\left( l,j,M\right) $ is the collection of the three
degenerated nonrelativistic quantum numbers and%
\begin{align*}
& \psi _{n,a}\left( r,\vartheta ,\varphi \right) \equiv \psi
_{n,l,j,M}\left( r,\vartheta ,\varphi \right) =R_{n,l}\left( r\right) \chi
_{j,M}\left( \vartheta ,\varphi \right) \,, \\
& \boldsymbol{\hat{\jmath}}^{2}\chi _{j,M}\left( \vartheta ,\varphi \right)
=\left( \mathbf{\hat{L}+}\boldsymbol{\sigma }/2\right) ^{2}\chi _{j,M}\left(
\vartheta ,\varphi \right) =j\left( j+1\right) \chi _{j,M}\left( \vartheta
,\varphi \right)
\end{align*}%
are the eigenfunctions of (\ref{1.26b}), $\hat{H}^{\mathrm{nr}}\psi
_{n,a}=E_{n}\psi _{n,a}$.

In order to obtain the $\theta $-modification of the $%
2S_{1/2},~2P_{1/2},~2P_{3/2}$ levels, we emphasize that $\hat{V}^{\theta }$
is a vectorial operator and, according to angular momentum selections rules,
allowed transitions are between levels with $\Delta M\equiv \left\vert
M-M^{\prime }\right\vert =0,1$ and $\Delta l\equiv \left\vert l-l^{\prime
}\right\vert =0$ \cite{Landau3}. The $6\times 6$ secular matrix $\Delta E^{%
\mathrm{nr}}$ has three degenerate eigenvalues represented as $\Delta E_{0}^{%
\mathrm{nr}}$ and $\Delta E_{\pm }^{\mathrm{nr}}$ and, restoring $\hbar $
and $c$, these corrections have the form

\begin{equation}
\Delta E_{0}^{\mathrm{nr}}=0~,\ \Delta E_{\pm }^{\mathrm{nr}}=-\frac{mc^{2}}{%
4}\left( \frac{Z^{2}\alpha ^{2}}{\lambdabar _{e}}\right) ^{2}\left( \frac{1}{%
24}\right) \left( \pm \left\vert \boldsymbol{\theta }\right\vert \right)
\,,\ \ \left\vert \boldsymbol{\theta }\right\vert =\sqrt{\theta ^{i}\theta
^{i}}\,,  \label{1-29}
\end{equation}%
where $\alpha =e^{2}/\hbar c$ is the fine structure constant, $\lambdabar
_{e}=\lambda _{e}/2\pi =\hbar /mc$ and $\lambda _{e}$ is the electron
Compton wavelength. The same results are obtained if we choose $\theta
_{1}=\theta _{2}=0$ and $\theta _{3}\neq 0$ (which can be done by a rotation
or a redefinition of coordinates). With this choice, $\left\vert \boldsymbol{%
\theta }\right\vert =\theta _{3}$, and the diagonal elements can be
calculated using the general formulae obtained in \cite{ChaiSheTur01}. Since
the goal of \cite{ChaiSheTur01} is the study of the transition \footnote{%
Rigorously this transition does not exist, as the perturbed level is
actually a superposition of the degenerated levels $2P_{1/2}$ and $2P_{3/2}$%
, and only the quantum number $M$ is well defined. But this consideration
does not change the numerical value of the transition and, consequently,
could be ignored in \cite{ChaiSheTur01}.} $2P_{1/2}\rightarrow 2S_{1/2}$, it
does not contain the expression for the nondiagonal elements of $\Delta
E_{bb^{\prime }}^{\mathrm{nr}}$. For these nondiagonal elements, we have%
\begin{align}
\Delta E_{bb^{\prime }}^{\mathrm{nr}}& =-\frac{Ze^{2}}{4}\int_{0}^{\infty
}dr\,r^{2}\int_{0}^{4\pi }d\Omega \left\{ \psi _{2,b}^{\dagger }\left(
r,\vartheta ,\varphi \right) \left[ \frac{\mathbf{\hat{L}}\cdot \boldsymbol{%
\theta }}{r^{3}}\right] \psi _{2,b^{\prime }}\left( r,\vartheta ,\varphi
\right) \right\}  \notag \\
& =-\frac{mc^{2}}{4}\left( \frac{Z^{2}\alpha ^{2}}{\lambdabar _{e}}\right)
^{2}\left( \frac{1}{24}\right) \left( \frac{\sqrt{2}}{3}\theta _{3}\right)
\,,\ \ b=\left( 1,\frac{1}{2},\pm \frac{1}{2}\right) \,,\ \ b^{\prime
}=\left( 1,\frac{3}{2},\pm \frac{1}{2}\right) \,.  \label{1-30}
\end{align}%
With this definition of coordinates, $\left[ \hat{J}_{3,}\hat{V}^{\theta }%
\right] =0$, therefore the eigenstates of $\hat{H}^{\theta ,\mathrm{nr}}=%
\hat{H}^{\mathrm{nr}}+\hat{V}^{\theta }$, in leading order, have the
magnetic quantum number $M$ well defined. This result allows us to conclude
that, in nonrelativistic case, the $2S-2P$ levels split in three other
levels, all of them \emph{twofold degenerate}, as illustrated in Figure 1.

In the relativistic theory, the $\theta $-modification of the energy levels
are obtained by computing the eigenvalues of the secular matrix $\Delta E^{%
\mathrm{rel}}$, characterized by the mean values of the operator $\hat{V}%
^{\theta }$ (\ref{1.2}), but now with respect to the Dirac spinors $\Psi
_{N,j,M,\zeta }$ (\ref{1.12}), with the same angular momentum selection
rules, i.e., $\Delta M\equiv \left\vert M-M^{\prime }\right\vert =0,1$ and $%
\Delta l\equiv \left\vert l-l^{\prime }\right\vert =0 $. The perturbation
operator $\hat{V}^{\theta }$ does not mix the small $F_{N,j,\zeta
}^{-}\left( r\right) $ and the big $F_{N,j,\zeta }^{+}\left( r\right) $\
components of the Dirac spinor (\ref{1.12}), so $\Delta E^{\mathrm{rel}}$
consists of two terms, which read:
\begin{align*}
\Delta E_{\mu \mu ^{\prime }}^{\mathrm{rel}}& =-\frac{Ze^{2}}{4}%
\int_{0}^{\infty }\frac{dr}{r}\left[ \left( F_{N,j,\zeta }^{+}\right) ^{\ast
}F_{N,j,\zeta ^{\prime }}^{+}\right] \int_{0}^{4\pi }d\Omega \left\{ \Omega
_{j,M,\zeta }^{\dagger }\left( \mathbf{\hat{L}}\cdot \boldsymbol{\theta }%
\right) \Omega _{j,M^{\prime },\zeta ^{\prime }}\right\} \\
& -\frac{Ze^{2}}{4}\int_{0}^{\infty }\frac{dr}{r}\left[ \left( F_{N,j,\zeta
}^{-}\right) ^{\ast }F_{N,j,\zeta ^{\prime }}^{-}\right] \int_{0}^{4\pi
}d\Omega \left\{ \Omega _{j,M,-\zeta }^{\dagger }\left( \mathbf{\hat{L}}%
\cdot \boldsymbol{\theta }\right) \Omega _{j,M^{\prime },-\zeta ^{\prime
}}\right\} \,,
\end{align*}%
where the label $\mu =\left( M,\zeta \right) $ is the set of two degenerated
relativistic quantum numbers. However, $F_{N,j,\zeta }^{-}\left( r\right) $
is approximately by $1/c$ smaller then $F_{N,j,\zeta }^{+}\left( r\right) $,
so we can neglect the terms proportional to small components, retaining only
the first term above.

We are interested in calculating the $\theta $-modifications of the
relativistic energy of the $2S$,$\ 2P$ levels. The nonrelativistic
degeneracy in the quantum number $j$ is naturally removed in the
Dirac theory, in such a way that the level $2P_{3/2}$ is lifted from
the still degenerate $2S_{1/2}$, $~2P_{1/2}$ levels. The space
noncommutativity leads to additional splittings and a separate
analysis for the $j=1/2$ and $j=3/2$ relativistic
energy levels is required. The elements of the secular matrix $\Delta E^{%
\mathrm{rel}}\left( j\right) $, in leading order, are
\begin{equation}
\Delta E_{MM^{\prime }}^{\mathrm{rel}}\left( j\right) =-\frac{Ze^{2}}{4}%
\int_{0}^{\infty }dr\,r^{2}\left\vert F_{N,j,\zeta }^{+}\right\vert
^{2}\int_{0}^{4\pi }d\Omega \left\{ \Omega _{j,M,\zeta }^{\dagger }\left[
\frac{\mathbf{\hat{L}}\cdot \boldsymbol{\theta }}{r^{3}}\right] \Omega
_{j,M^{\prime },\zeta }\right\} \,,  \label{1-31}
\end{equation}%
whose results, for each level, follow below.

\subsection{Relativistic NC correction for the $2P_{1/2}$ level}

The $\theta $-correction for the $2P_{1/2}$ level $\left(
N=1\,,~j=1/2\,,~\zeta =1\,,~M=\pm \frac{1}{2}\right) $ follows from (\ref%
{1-31}):%
\begin{align}
& \Delta E_{MM^{\prime }}^{\mathrm{rel}}\left( 1/2\right) =-\frac{Ze^{2}}{4}%
\varrho _{1/2}\Theta _{MM^{\prime }}^{\mathrm{rel}}\,,  \notag \\
& \varrho _{1/2}=\int_{0}^{\infty }\frac{dr}{r}\left\vert F_{1,\frac{1}{2}%
,1}^{+}\left( r\right) \right\vert ^{2}\,,\ \ \Theta _{MM^{\prime }}^{%
\mathrm{rel}}=\int_{0}^{4\pi }d\Omega \left[ \Omega _{\frac{1}{2},M,1}^{\ast
}\left( \mathbf{\hat{L}}\cdot \boldsymbol{\theta }\right) \Omega _{\frac{1}{2%
},M^{\prime },1}\right] \,.  \label{1-32}
\end{align}%
Expressing the hypergeometric confluent functions (\ref{1.13}) in terms of
exponentials and polynomials, we obtain%
\begin{align}
& \varrho _{1/2}=\left( 2\lambda _{1}\right) ^{3}\left[ \frac{\beta
_{1}\left( \lambda _{1}^{+}\right) ^{2}}{4\eta _{1}\left( \eta _{1}-1\right)
\left( \beta _{1}-1\right) \left( \beta _{1}-2\right) \left( \beta
_{1}-3\right) }\right] \left\{ \left( \eta _{1}-2\right) ^{2}\right.   \notag
\\
& -\left. \left[ \frac{2\left( \eta _{1}-2\right) \left( \eta _{1}-1\right)
\left( \beta _{1}-3\right) }{\beta _{1}}\right] +\left( \frac{\eta _{1}-1}{%
\beta _{1}}\right) ^{2}\left( \beta _{1}-2\right) \left( \beta _{1}-3\right)
\right\} \,,  \label{1.16} \\
& \eta _{1}=Ze^{2}\frac{m}{\lambda _{1}}\,,\ \,\gamma _{1}=\sqrt{1-\left(
Ze^{2}\right) ^{2}}\,,\ \,\beta _{1}=2\gamma _{1}+1\,,  \notag
\end{align}%
with $\lambda _{1}^{+}$, $\lambda _{1}$ given in (\ref{1.13}). The matrix $%
\Theta ^{\mathrm{rel}}$, whose elements are $\Theta _{MM^{\prime }}^{\mathrm{%
rel}}$, reads%
\begin{equation}
\Theta ^{\mathrm{rel}}=\frac{2}{3}\left(
\begin{array}{cc}
-\theta _{3} & \theta _{+} \\
\theta _{-} & \theta _{3}%
\end{array}%
\right) \,,\ \ \theta _{\pm }=\theta _{1}\pm i\theta _{2}\,,  \label{1.7}
\end{equation}%
and its eigenvalues, denoted by $\Lambda _{j}^{\pm }$, are%
\begin{equation}
\Lambda _{1/2}^{\pm }=\pm \frac{2}{3}\left\vert \boldsymbol{\theta }%
\right\vert \,,\ \ \left\vert \boldsymbol{\theta }\right\vert =\sqrt{\theta
^{i}\theta ^{i}}\,.  \label{1-35}
\end{equation}

The $\theta $-corrections to the relativistic energy of the $2P_{1/2}$
level, $\Delta E_{\pm }^{\mathrm{rel}}\left( j\right) $, restoring $\hbar $
and $c$, reads

\begin{align}
& \Delta E_{\pm }^{\mathrm{rel}}\left( 1/2\right) =-\frac{Ze^{2}}{4}\varrho
_{1/2}\Lambda _{1/2}^{\pm }  \notag \\
& =-mc^{2}\left( \frac{Z^{2}\alpha ^{2}}{\lambdabar _{e}}\right) ^{2}\left(
\pm \frac{\left\vert \boldsymbol{\theta }\right\vert }{3}\right) \left\{
\left( \frac{\left( \lambda _{1}^{+}\right) ^{2}\beta _{1}}{\left( \beta
_{1}-1\right) \left( \beta _{1}-2\right) \left( \beta _{1}-3\right) }\right)
\right.  \notag \\
& \times \left( \frac{1}{\eta _{1}^{4}\left( \eta _{1}-1\right) }\right) %
\left[ \left( \eta _{1}-2\right) ^{2}-\left( \frac{2\left( \eta
_{1}-2\right) \left( \eta _{1}-1\right) \left( \beta _{1}-3\right) }{\beta
_{1}}\right) \right.  \notag \\
& +\left. \left. \left( \frac{\eta _{1}-1}{\beta _{1}}\right) ^{2}\left(
\beta _{1}-2\right) \left( \beta _{1}-3\right) \right] \right\} \,.
\label{1-36}
\end{align}

Therefore there is a splitting of this level in such way that the originally
degenerate level $2S_{1/2}\,,~2P_{1/2}$ splits in three sublevels. Similarly
to the nonrelativistic case, the same results can be obtained by choosing a
reference frame where $\theta_{1}=\theta_{2}=0$ and $\theta_{3}\neq0$. The
advantage of this frame is that $\left[ \hat{J}_{3},\hat {V}^{\theta}\right]
=0$ and therefore, the new states have the magnetic quantum number $M$ well
defined. In this reference frame, the splitting of the level $%
2S_{1/2}\,,~2P_{1/2}$ in the three levels $%
2P_{1/2}^{+1/2},~2S_{1/2},~2P_{1/2}^{-1/2}$ is illustrated in Figure 1.

\subsection{Relativistic NC correction for the $2P_{3/2}$ level}

According to (\ref{1-31}), the $\theta $-correction for the $2P_{3/2}$ level
($N=0$, $j=3/2$, $\zeta =-1$, $M=\pm 1/2,\,\pm 3/2 $) is%
\begin{equation}
\Delta E_{MM^{\prime }}^{\mathrm{rel}}\left( 3/2\right) =-\frac{Ze^{2}}{4}%
\varrho _{3/2}\Theta _{MM^{\prime }}^{\mathrm{rel}}\,,  \label{1-37}
\end{equation}%
where the radial integral is%
\begin{align}
& \varrho _{3/2}=\int_{0}^{\infty }\frac{dr}{r}\left\vert F_{0,\frac{3}{2}%
,-1}^{+}\left( r\right) \right\vert ^{2}=4\left( \lambda _{0}\right) ^{3}%
\left[ \frac{\left( \lambda _{0}^{+}\right) ^{2}}{\left( \beta _{2}-1\right)
(\beta _{2}-2)(\beta _{2}-3)}\right] \,,  \label{1-38} \\
& \gamma _{2}=\sqrt{4-\left( Ze^{2}\right) ^{2}}\,,\ \ \beta _{2}=2\gamma
_{2}+1\,,  \notag
\end{align}%
with $\lambda _{0}^{+}$, $\lambda _{0}$ given in (\ref{1.13}). Compared to
the previous case, the matrix $\Theta ^{\mathrm{rel}}$ is now a $4 \times 4$%
-matrix, whose elements are%
\begin{equation}
\Theta _{MM^{\prime }}^{\mathrm{rel}}=\int_{0}^{4\pi }d\Omega \left[ \Omega
_{\frac{3}{2},M,-1}^{\ast }\left( \mathbf{\hat{L}}\cdot \boldsymbol{\theta }%
\right) \Omega _{\frac{3}{2},M^{\prime },-1}\right] \,,  \label{1-39}
\end{equation}%
and explicitly is given by%
\begin{equation}
\Theta ^{\mathrm{rel}}=\left(
\begin{array}{cccc}
-\theta _{3} & \frac{1}{\sqrt{3}}\theta _{+} & 0 & 0 \\
\frac{1}{\sqrt{3}}\theta _{-} & -\frac{1}{3}\theta _{3} & \frac{2}{3}\theta
_{+} & 0 \\
0 & \frac{2}{3}\theta _{-} & \frac{1}{3}\theta _{3} & \frac{1}{\sqrt{3}}%
\theta _{+} \\
0 & 0 & \frac{1}{\sqrt{3}}\theta _{-} & \theta _{3}%
\end{array}%
\right) \,,\ \ \theta _{\pm }=\theta _{1}\pm i\theta _{2}\,.  \label{1.8}
\end{equation}%
Its four nondegenerate eigenvalues, $\Lambda _{j}^{\pm }$ and $\Lambda
_{j}^{\pm \prime }$, are%
\begin{equation}
\Lambda _{3/2}^{\pm }=\pm \left\vert \boldsymbol{\theta }\right\vert \,,\ \
\Lambda _{3/2}^{\pm \prime }=\frac{\Lambda _{3/2}^{\pm }}{3}\,,\ \
\left\vert \boldsymbol{\theta }\right\vert =\sqrt{\theta ^{i}\theta ^{i}}\,.
\label{1-40}
\end{equation}%
The $\theta $-corrections for the $2P_{3/2}$ level, in dimension of energy
(restoring $\hbar $ and $c$), have the form%
\begin{align}
& \Delta E^{\mathrm{rel}}\left( 3/2\right) ^{\pm }=-\frac{Ze^{2}}{4}\varrho
_{3/2}\Lambda _{3/2}^{\pm }~,\,\Delta E^{\mathrm{rel}}\left( 3/2\right)
^{\pm \prime }=\frac{\Delta E^{\mathrm{rel}}\left( 3/2\right) ^{\pm }}{3}~,
\notag \\
& \Delta E^{\mathrm{rel}}\left( 3/2\right) ^{\pm }=-\frac{mc^{2}}{4}\left(
\frac{Z^{2}\alpha ^{2}}{\lambdabar _{e}}\right) ^{2}\left( \pm \left\vert
\boldsymbol{\theta }\right\vert \right) \frac{\left( \lambda _{0}^{+}\right)
^{2}}{2\left( \beta _{2}-1\right) (\beta _{2}-2)(\beta _{2}-3)}~.
\label{1-42}
\end{align}%
Choosing again the reference frame where $\theta _{1}=\theta _{2}=0$ and $%
\theta _{3}\neq 0,$ we have $\left[ \hat{J}_{3},\hat{V}^{\theta }\right] =0$
and the level $2P_{3/2}$ splits in the four levels $2P_{3/2}^{+1/2}%
\,,~2P_{3/2}^{-1/2},~2P_{3/2}^{+3/2},~2P_{3/2}^{-3/2}$, as illustrated in
Figure 1.

The energy levels and their modifications due to presence of space
noncommutativity, for
the relativistic and nonrelativistic cases, are shown\footnote{%
On this figure, the energy splits indicated by the dashed-point arrow comes
from the usual Dirac equation, so it does not depends on $\theta$ and this
split can not be compared with the others $\theta$-modified levels.} in
Figure 1.


\begin{center}
\includegraphics[
height=2.884in, width=6.2109in
]{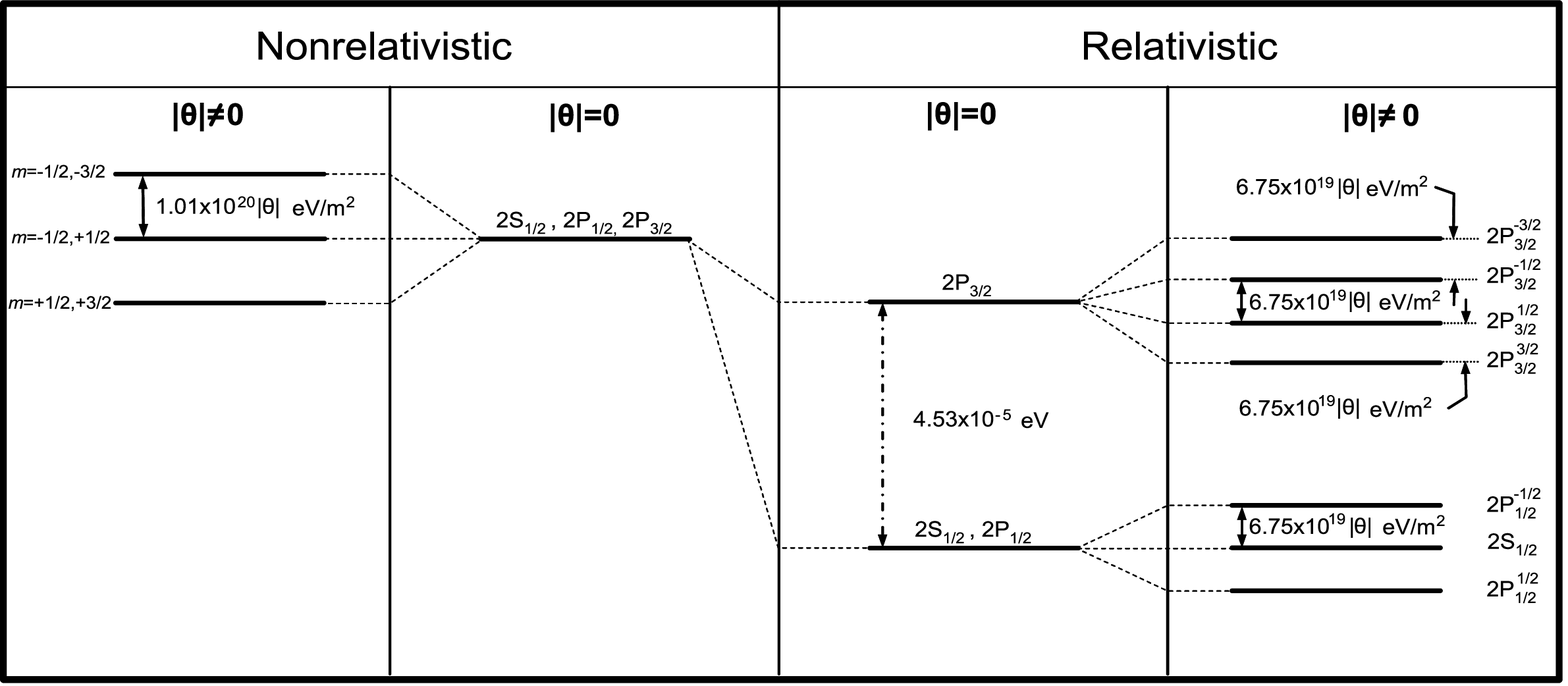}\\[0pt]\end{center}
\textbf{Figure 1} -- Splittings for relativistic and nonrelativistic
energy levels due to space noncommutativity. All numerical values of
the coefficients of the first-order $\theta$-corrections are in
units of eV/m$^2$. \label{Figure1}

\vskip 0.3cm


We conclude that, from the point of view of the $\theta $-modified
relativistic Dirac theory there is an additional, in contrast to the
nonrelativistic case, splitting of some degenerate levels and there
appear new transition channels. In particular, in the presence of
the space noncommutativity, the degenerate levels
$2S_{1/2},2P_{1/2}$\ split into three nondegerenerate sublevels and
the level $2P_{3/2}$ splits into four nondegenerate sublevels, and
the transition $2P_{1/2}^{\pm 1/2}\longrightarrow 2S_{1/2}$ is
possible. Except for the spherically symmetric levels $2S_{1/2}^{\pm
1/2}$,\ these results show that in the noncommutative relativistic
theory degeneracy is completely removed. The obtained results show
explicitly the dependence of the energy levels of the hydrogen atom
on the noncommutative parameter. Once these energy levels can be
measured experimentally with a high accuracy, the actual
spectroscopy data can be used to impose bounds on the value of the
noncommutativity parameter $\theta $.

\section*{Acknowledgements}

T.C.A. thanks CNPq; M.C.B. thanks FAPESP; D.M.G. thanks FAPESP and
CNPq for permanent support. The support of the Academy of Finland
under the Projects no. 121720 and 127626 is greatly acknowledged.

\end{document}